# Multi-agent based modeling for investigating excess heat utilization from electrolyzer production to district heating network

Christensen, Kristoffer; Jørgensen, Bo Nørregaard; Ma, Zheng Grace

# Multi-agent based modeling for investigating excess heat utilization from electrolyzer production to district heating network

Kristoffer Christensen[1][0000-0003-2417-338X], Bo Nørregaard Jørgensen[1][0000-0001-5678-6602] and Zheng Grace Ma[1][0000-0002-9134-1032]

[1] SDU Center for Energy Informatics, Maersk Mc-Kinney Moeller Institute, The Faculty of Engineering, University of Southern Denmark, Odense, Denmark
zma@mmmi.sdu.dk

**Abstract.** Power-to-Hydrogen is crucial for the renewable energy transition, yet existing literature lacks business models for the significant excess heat it generates. This study addresses this by evaluating three models for selling electrolyzer-generated heat to district heating grids: constant, flexible, and renewable-source hydrogen production, with and without heat sales. Using agent-based modeling and multi-criteria decision-making methods (VIKOR, TOPSIS, PROMETHEE), it finds that selling excess heat can cut hydrogen production costs by 5.6%. The optimal model operates flexibly with electricity spot prices, includes heat sales, and maintains a hydrogen price of 3.3 EUR/kg. Environmentally, hydrogen production from grid electricity could emit up to 13,783.8 tons of $CO_2$ over four years from 2023. The best economic and environmental model uses renewable sources and sells heat at 3.5 EUR/kg.

## 1       Introduction

The extensive deployment of renewable energy sources in the green transition is facilitated by sector coupling, which aims to reduce emissions across various sectors. For example, green electricity from wind is utilized for heating via heat pumps or for powering electric vehicles in transportation. However, options for electrifying heavy transportation, such as cargo ships and airplanes, are limited due to the constraints of battery technology [1]. To bridge this gap, the Power-to-X (P2X) concept is introduced, enabling broad sector coupling through the production of renewable hydrogen. This hydrogen can then be used directly or converted into other fuels like methanol and ammonia.

This paper focuses on the potential of using excess heat from electrolyzers in the P2X value chain for district heating. Specifically, it examines the Power-to-Hydrogen ecosystem, excluding other segments of the value chain post-hydrogen production. The potential for heat utilization through three business models for electrolyzer operation is explored. These models are simulated in an agent-based framework and evaluated based on their Levelized Cost of Hydrogen (LCoH), both with and without the sale of excess heat. To determine the optimal operation, three Multi-Criteria Decision-Making (MCDM) methods are applied.

The structure of this paper is as follows: it begins with a background on the utilization of excess heat from electrolyzers, followed by an explanation of the methodology used. The case study introduction sets the stage for scenarios and simulation experiments, which detail the conducted simulations and present the findings in the results and discussion section. The paper concludes with a summary of the key findings.

## 2      Background

The majority of the literature considers the Power-to-Hydrogen and Heat (P2HH) in a system, including a Combined Heat and Power (CHP) plant. CHPs' heat production is constrained due to the co-production of electricity. Here, the excess heat utilization from P2X facilities can provide more flexibility to the CHP plant [2-4].

The characteristics of P2HH found in the literature are the use of the electrolyzer's excess heat via a heat exchanger in the district heating network. The electrolyzer technologies used are Alkaline (operating temperature of 60–90 °C) or Proton Exchange Membrane (PEM) (operating temperature of 50–80 °C). The electrolyzer is modeled using the Temperature, Heat, and Hydrogen (T-H-H) relationship for modeling the electrolyzer. The Mixed Integer Linear Programming is commonly used to optimize the operation.

The available relatively high temperatures from the electrolysis in the Proton Exchange Membrane or Polymer Electrolyte Membrane (PEM) electrolyzer, and Alkaline electrolyzer (most common and mature electrolyzers) are high enough for supplying directly to the district heating network in Denmark (feed temperatures of ~60 °C [5]) through a heat exchanger. This means that there is no need for upgrading the temperature, meaning solutions concerning industrial excess heat recovery solutions are not relevant [6].

Currently, 96% of hydrogen in Europe comes from natural gas, and there is no specific market structure for green hydrogen [7]. Therefore, green hydrogen still primarily integrates into the larger gas market, often through conventional supply chains that might include on-site production or pre-agreed sales. From a national perspective, e.g., in Denmark, heat from the hydrogen production process has been accounted for within the P2X ecosystem. However, no literature has discussed business models for the utilization of excess heat, but rather for the electrolyzer operations through the classical business model of cross-commodity arbitrage trading (between the electricity market and the hydrogen market), as well as the provision of flexibility at control reserve markets [8, 9]. Three business models are identified based on the market search:

The most prevalent business model adopted by most Danish companies referred to as "Constant Hydrogen Production", is largely shaped by the prevailing conditions in the hydrogen market. This approach signifies a prominent pattern of consistent hydrogen generation irrespective of market fluctuations and demands. It reflects the traditional way of producing hydrogen from natural gas or methan (referred to as grey hydrogen) [10], which is assumed to be produced based on demand and hydrogen purchase agreements [11].

A second business model, known as "Flexible Production Following Electricity Prices" also known as cross-commodity arbitrage trading, mirrors the current hourly electricity pricing scheme for Danish electricity consumers, including households. This model incorporates dynamic pricing, where the cost of production fluctuates in line with the prevailing electricity prices [8, 12, 13]. It is assumed that the hydrogen purchase agreements account for this operation type.

The third business model, termed "Hydrogen Production from Own Renewable Energy Sources", encapsulates the pilot projects facilitated by Danish DSO – Energinet [14]. This model ensures the production of green hydrogen and potentially allows to access cheaper electricity prices due to its associated renewable energy generations.

## 3     Methodology

New business models for large-scale electrolyzers are expected to evolve due to maturing digital energy solutions that are expected to improve energy efficiency and enable energy flexibility, which are key value propositions of the P2X concept [15]. To ensure that this research accounts for future changes in the business ecosystem, it will follow the agent-based simulation framework for evaluating energy flexibility solutions and adoption strategies proposed in [16]. In the framework, the targeted business ecosystems are identified and translated into multi-agent based models that capture all ecosystem dynamics, and the associated solutions can be evaluated.

To navigate the complexity of the P2X business ecosystem, this study employs agent-based modeling, capturing ecosystem dynamics by representing actors as agents with defined roles and interactions modeled through Java interfaces [17, 18]. This model is developed using the multi-method simulation tool AnyLogic [19]. For simplicity, the model focuses on Power-to-Hydrogen, but it is designed to allow easy inclusion of additional agents and interactions to reflect a broader ecosystem [20, 21]. Key agents include:
- Electrolyzer: Processes water and electricity to produce hydrogen and excess heat.
- Transmission System Operator (TSO): Manages day-ahead electricity prices and provides grid tariffs at the transmission level, ignoring the market dynamics in the simulation.
- Heat Exchanger: Transfers heat between the electrolyzer's cooling system and the district heating grid, accounting for efficiency. It may be replaced by heat pumps if needed.
- District Heating Grid: Acts as a simple receiver of all heat produced by the electrolyzer without further detail.
- District Heating Operator: Adds pricing to the heat supplied by the electrolyzer.

- P2X Company: Handles operational logic within the electrolyzer, calculating operation costs, revenue, and profit.
- Electricity Grid: Connects directly to the transmission level and is considered a black box in this scenario, eventually to be detailed further.
- Hydrogen Customer: Engages in contractual arrangements with the P2X Company to purchase hydrogen.
- Hydrogen Infrastructure: Represents the endpoint for hydrogen flow, consisting of storage and pipeline flow.
- Water Treatment Plant: Provides the electrolyzer with clean water, potentially expanding to include resource usage data based on raw water sources.
- Wind Turbine: Supplies the electrolyzer with electricity from wind, with cost metrics compared against market prices.
- Photovoltaics: Converts solar radiation into electricity, important for hydrogen production from renewable resources.

The simulation calculates variable costs such as electricity and water, and revenues from selling heat and hydrogen to determine the LCoH. The operation hours, influenced by the operation strategy, affect the LCoH. The most effective operation is evaluated with MCDM methods (PROMETHEE, TOPSIS, and VIKOR) [22]. These methods, programmed using the "pymcdm" Python library [23], assess simulation outcomes with both equal and entropy weighting to objectively weigh decision criteria.

## 4    Case study

The case study used in this paper is an industrial park, Greenlab Skive, Denmark [24], which among other facilities consists of a large-scale 12 MW electrolysis plant together with 80 MW of renewable energy resources and is located close to an existing district heating grid. The electrolyzer is Alkaline as it is the most common electrolyzer type used in Denmark. Table 1 shows the parameters used in the simulation. The prices shown in the table are assumed to be the same in the future, meaning they are used in the future years in the simulation. For example, electricity prices are looped throughout the simulation.

Table 1. Simulation model parameter input.

| Parameter | Value | Ref. |
|---|---|---|
| Electricity spot prices | *Dataset from 2015-2018 | [25] |
| Grid tariff | 13.4 EUR/MWh | [26] |
| Price for district heating | 20.1 and 26.8 EUR/MWh for 70°C in summer and winter respectively (price cap of 44.9 EUR/MWh [27]) | [28] |
| Deionized water consumption | 10 L/kg H2 | [29] |
| Price for water used in electrolyzer | 0.85 EUR Cents/L (calculated based on values from source) | [28] |
| Energy balance for electrolysis | H2: 66.5%, Recoverable heat loss: 16.4% for Alkaline Electrolyzer | [30] |

| Parameter | Value | Ref. |
|---|---|---|
| Hydrogen density | Lower heating value (the one used in the energy balance) is 120 MJ/kg H2 = <u>33.33 kWh/kg H2</u> | [30] |
| Electrolyzer Capacity | 12 $MW_E$ | [31] |
| Electrolyzer lifetime | 20 years | [28] |
| Investment cost (CapEx) | 10 million per $MW_E$ | [28] |
| Operation and maintenance cost | 3% of CapEx per year | [28] |
| Hydrogen market price | **Price EUR/kg of hydrogen: 1.5 (grey), 2 (blue), 2.7 (used in [28] case study) 3.5 (break-even in [28]). | [28] |

*Electricity prices from 2019 to now is not estimated to be representable for future prices due to COVID-19 and the Russian-Ukranian war.

**The types of hydrogen are defined as follows: grey hydrogen is produced primarily from natural gas, while blue hydrogen is similar but includes carbon capture and storage [10].

## 5    Scenarios and simulation experiments

The simulation scenarios incorporate three business models: Scenario 1 involves constant hydrogen production; Scenario 2 adapts production based on electricity prices; Scenario 3 generates hydrogen using renewable energy sources. Additionally, all scenarios test selling excess heat directly to the district heating grid using a heat exchanger.

Simulations are conducted for each hydrogen market price listed in Table 1, focusing on experiments unrelated to determining hydrogen prices. Simulations span from 2023 to at least 2026, utilizing available electricity price data. The simulations conclude at the point of Return on Investment (RoI) determination, or by 2050 if no RoI is established earlier. Table 2 details the setup for these experiments, e.g., experiment 1.1 is Scenario 1 without excess heat sales, and 1.2 includes it.

For Scenario 3, parameters in Table 3 calculate the levelized cost of energy for the onsite renewable energy park, thereby determining the LCoH. The MCDM evaluation uses Key Performance Indicators (KPIs) such as operation hours, LCoH, hydrogen, and heat production, revenues from hydrogen and heat, $CO_2$ emissions, and profit to compare experiments.

**Table 2.** Simulation experiments.

| Experiment number | Electricity source | Electricity price scheme | Electrolyzer operation mode | Price for hydrogen [EUR/kg] | Price for utilized excess heat [EUR/MWh] |
|---|---|---|---|---|---|
| 1.1 | From grid | Spot price + grid tariff | Constant production | 1.5, 2, 2.7, and 3.5 | No heat sold |
| 1.2 | | | | | 20.1 and 26.8* |
| 2.1 | | | Flexible production | | No heat sold |
| 2.2 | | | | | 20.1 and 26.8* |
| 3.1 | renewable energy park | | Constant production | | No heat sold |
| 3.2 | | | | | 20.1 and 26.8* |

*summer and winter, respectively

**Table 3.** Financial data on large-scale on-shore wind turbines and utility-scale ground-mounted photovoltaics [32]

|  | Unit | Wind turbines | photovoltaic system |
|---|---|---|---|
| CapEx per installed kW | EUR/kW | 1,126 | 452.4 |
| Variable O&M | EUR Cents/kWh | 0.15 | 0 |
| Fixed O&M | EUR/kW/year | 14.1 | 9 |
| Discount rate | % | 3.5 | 3.5 |
| Expected lifetime | Years | 27 | 35 (12.5)* |

*12.5 years for the inverter, which has an investment cost of 20.1 EUR/kW.

## 6     Results and discussions

### 6.1    Scenario results

All experimental results are outlined in Table 4. For scenario 1, experiments 1.1 and 1.2 show LCoH of 3.06 EUR/kg and 2.86 EUR/kg, respectively. The results indicate that hydrogen prices of 1.5 and 2 EUR/kg are not profitable, although prices at 2.7 EUR/kg do yield short-term profits by excluding capital and maintenance costs from calculations. The operations, averaging 8,709.75 hours annually over four years, produce 2,085.17 tons of hydrogen and approximately 14 tons of $CO_2$ annually.

**Table 4.** Key Performance Indicators for each experiment.

| Experiment | Hydrogen price | LCoH [EUR/kg] | Avg. yearly profit [Thousand EUR] | Avg. yearly operation hours | Avg. yearly produced hydrogen [ton] | Avg. yearly $CO_2$ emissions [ton] | Return on Investment* [Years] |
|---|---|---|---|---|---|---|---|
| 1.1 |  | 3.1 | -1,728.9 | 8,709.75 | 2,085.17 | 13,783.76 | - |
| 1.2 |  | 2.9 | -1,328.4 | 8,709.75 | 2,085.17 | 13,783.76 | - |
| 2.1 | 1.5 | 487.1 | 4.1 | 12.50 | 3.23 | 20.96 | - |
| 2.2 |  | 178.8 | 10.2 | 37.50 | 8.80 | 55.14 | - |
| 3.1 |  | 5.8 | -1,032.2 | 2,515.00 | 603.77 | 0 | - |
| 3.2 |  | 5.6 | -917.2 | 2,515.00 | 603.77 | 0 | - |
| 1.1 |  | 3.1 | -610.8 | 8,709.75 | 2,085.17 | 13,783.76 | - |
| 1.2 |  | 2.9 | -210.3 | 8,709.75 | 2,085.17 | 13,783.76 | - |
| 2.1 | 2 | 81.1 | 25.4 | 80.00 | 19.57 | 121.10 | - |
| 2.2 |  | 52.8 | 40.7 | 124.50 | 30.16 | 178.07 | - |
| 3.1 |  | 5.8 | -708.4 | 2,515.00 | 603.77 | 0 | - |
| 3.2 |  | 5.6 | -593.4 | 2,515.00 | 603.77 | 0 | - |
| 1.1 |  | 3.1 | 786.7 | 8,709.75 | 2,085.17 | 13,783.76 | 18.70 |
| 1.2 |  | 2.9 | 1,187.2 | 8,709.75 | 2,085.17 | 13,783.76 | 12.92 |
| 2.1 | 2.7 | 7.5 | 324.9 | 1,085.50 | 260.23 | 1,170.00 | - |
| 2.2 |  | 3.5 | 905.1 | 3,391.50 | 812.10 | 4,255.63 | - |
| 3.1 |  | 5.8 | -303.8 | 2,515.00 | 603.77 | 0 | - |
| 3.2 |  | 5.6 | -188.8 | 2,515.00 | 603.77 | 0 | - |
| 1.1 | 3.5 | 3.1 | 2,463.8 | 8,709.75 | 2,085.17 | 13,783.76 | 6.00 |

| Experiment | Hydrogen price | LCoH [EUR/kg] | Avg. yearly profit [Thousand EUR] | Avg. yearly operation hours | Avg. yearly produced hydrogen [ton] | Avg. yearly CO2 emissions [ton] | Return on Investment* [Years] |
|---|---|---|---|---|---|---|---|
| 1.2 |  | 2.9 | 2,864.3 | 8,709.75 | 2,085.17 | 13,783.76 | 5.24 |
| 2.1 |  | 3.0 | 2,474.8 | 7,598.50 | 1,818.66 | 11,553.52 | 5.99 |
| 2.2 |  | 2.8 | 2,889.4 | 8,401.75 | 2,012.34 | 13,150.66 | 5.21 |
| 3.1 |  | 5.8 | 181.8 | 2,515.00 | 603.77 | 0 | - |
| 3.2 |  | 5.6 | 296.8 | 2,515.00 | 603.77 | 0 | - |

*If the cell is empty no Return on Investment time was found within the simulations years (27 years).

The impact of selling excess heat on the Return on Investment (RoI) is significant, decreasing the RoI period by about six years at a hydrogen price of 2.7 EUR/kg. The smaller impact at 3.5 EUR/kg is due to the constant prices of excess heat, which diminishes its relative contribution to profits.

In scenario 2 (experiments 2.1 and 2.2), the business model employs a flexible production strategy that adjusts to fluctuating electricity prices. Hydrogen prices significantly influence the results due to the management of flexible operations. At the start of the year, higher electricity prices result in a gradual decrease in operational hours, as it is too expensive to produce hydrogen. This demands lower electricity prices in the future production schedule to account for the increasing idle time. This operational model directly impacts the LCoH, with reduced hydrogen production leading to higher LCoH. Notably, at a hydrogen price of 2.7 EUR/kg, selling excess heat substantially lowers the LCoH. Comparison with a higher price of 3.5 EUR/kg demonstrates that flexible operations can enhance profitability without significantly affecting hydrogen output.

Additionally, this paper applies an evolutionary optimization algorithm [33] in scenario 2 to determine the most competitive fixed hydrogen price that minimizes LCoH, ensuring production costs remain low without compromising profitability. It's crucial to recognize how hydrogen pricing affects the electrolyzer's operation. High spot prices can still yield profits, but excessively high prices lead to costlier hydrogen production. Ideally, the operation should occur at price points that maximize profitability and maintain low LCoH.

Results in Table 5 reveal that the optimal hydrogen production cost, adjusted for electricity prices between 2015 and 2018, ranges from 3.50 EUR/kg without excess heat to 3.30 EUR/kg with excess heat usage. Achieving these rates involves reducing hydrogen production by approximately 12% from 2,085.17 tons to about 1,838 tons, leveraging the flexibility of the electrolyzer. This adjustment, alongside selling excess heat to the district heating grid, can decrease hydrogen costs by an additional 5.6%.

The established optimal fixed hydrogen price of 2.79 EUR/kg for 7,682 operational hours annually does not limit the potential to sell hydrogen at higher prices but suggests following this pricing strategy to achieve the lowest production costs.

Table 5. Optimization results of minimizing the LCoH.

| KPI | Excluding the sale of excess heat | Including the sale of excess heat |
|---|---|---|
| Optimal hydrogen price [EUR/kg H2] | 3.498 | 3.301 |
| Lowest LCoH [EUR/kg H2] | 2.987 | 2.795 |
| Avg. yearly hydrogen production [ton] | 1,849.60 | 1,838.53 |
| Avg. yearly operation hours | 7,724.5 | 7,682.5 |

In scenario 3 (experiments 3.1 and 3.2), the electrolyzer operates on renewable energy, with profits being influenced directly by hydrogen prices due to the dependency on renewable output levels. The LCoH reaches 5.8 and 5.6 EUR/kg without and with the sale of excess heat, respectively, as recorded in Table 4. The operation drastically cuts operation hours by 71% compared to scenario 1, reflecting the renewable energy park's insufficiency in maximizing electrolyzer capacity. This is further underscored by the average wind speed of 3.9 m/s and solar radiation of 116.68 W/m², with wind turbines and PV systems requiring 13 m/s and 1000 W/m², respectively, to reach peak production [34] [35].

### 6.2    Multi-criteria decision-making results

The LCoH when compared to market rates, as shown in Table 4, indicates that the green hydrogen produced falls within the competitive range of 4-6 EUR/kg, distinguishing it from the less expensive grey and blue hydrogen options [36]. KPIs such as hydrogen price, LCoH, profit, hydrogen production, and emissions are used as decision criteria in the MCDM evaluation, utilizing both Equal and Entropy weightings (Table 6 and Table 7). The MCDM rankings, displayed in Fig. 1, identify scenario 3.2 with a hydrogen price of 3.5 EUR/kg as the most favorable, emphasizing zero emissions and profit, and underscoring the importance of environmental considerations in the evaluations.

Table 6. Weighting values.

| Decision criteria | Equal weight | Entropy weight |
|---|---|---|
| Hydrogen Price | 0.2 | 0.00583391 |
| LCoH | 0.2 | 0.19442628 |
| Avg. yearly profit | 0.2 | 0.37966943 |
| Avg. yearly produced hydrogen | 0.2 | 0.04040095 |
| Avg. yearly $CO_2$ emissions | 0.2 | 0.37966943 |

The average rankings across all experiments highlight experiment 3.2 with a hydrogen price of 1.5 EUR/kg as the optimal choice under equal weighting. Differences between the lowest and highest hydrogen prices in the same experiment predominantly affect profit rather than hydrogen price. The higher significance of profit changes, due to normalization across all scenarios rather than just scenario 3, makes hydrogen price more impactful in the evaluation. This observation suggests that equal weighting might not be appropriate unless hydrogen price is specifically considered. Consequently, an alternative weighting method has been applied.

Table 7. MCDM evaluation ranking with Entropy weight.

| Alternative | Experiment number (hydrogen price) | Average ranking of all three methods (Equal weighting) | Average ranking of all three methods (Entropy weighting) |
| --- | --- | --- | --- |
| A1 | 1.1 (1.5) | 15.3 | 17.3 |
| A2 | 1.2 (1.5) | 8.7 | 23.0 |
| A3 | 2.1 (1.5) | 23.0 | 15.3 |
| A4 | 2.2 (1.5) | 14.7 | 11.7 |
| A5 | 3.1 (1.5) | 6.7 | 15.0 |
| A6 | 3.2 (1.5) | **4.0** | 13.0 |
| A7 | 1.1 (2) | 14.7 | 22.0 |
| A8 | 1.2 (2) | 7.0 | 21.0 |
| A9 | 2.1 (2) | 18.3 | 9.7 |
| A10 | 2.2 (2) | 13.7 | 8.7 |
| A11 | 3.1 (2) | 8.7 | 12.3 |
| A12 | 3.2 (2) | 5.7 | 10.0 |
| A13 | 1.1 (2.7) | 13.7 | 17.3 |
| A14 | 1.2 (2.7) | 8.0 | 15.0 |
| A15 | 2.1 (2.7) | 16.0 | 6.3 |
| A16 | 2.2 (2.7) | 4.7 | 3.7 |
| A17 | 3.1 (2.7) | 11.3 | 9.7 |
| A18 | 3.2 (2.7) | 9.0 | 7.3 |
| A19 | 1.1 (3.5) | 17.3 | 15.0 |
| A20 | 1.2 (3.5) | 13.0 | 12.7 |
| A21 | 2.1 (3.5) | 14.7 | 11.0 |
| A22 | 2.2 (3.5) | 10.3 | 11.0 |
| A23 | 3.1 (3.5) | 21.3 | 3.7 |
| A24 | 3.2 (3.5) | 19.7 | **1.7** |

In the entropy-weighted MCDM evaluation, experiment 3.2 at 3.5 EUR/kg emerges as the superior option. The VIKOR, TOPSIS, and PROMETHEE methods show comparable rankings, with significant variations only in PROMETHEE. The entropy weightings reveal a strong emphasis on $CO_2$ emissions and profit, particularly benefiting scenario 3 where emissions are zero. Given the shift towards green hydrogen production via electrolyzers rather than traditional methods reliant on natural gas, high emissions weighting aligns with stakeholder interests, making the entropy weighting method an appropriate choice for these evaluations.

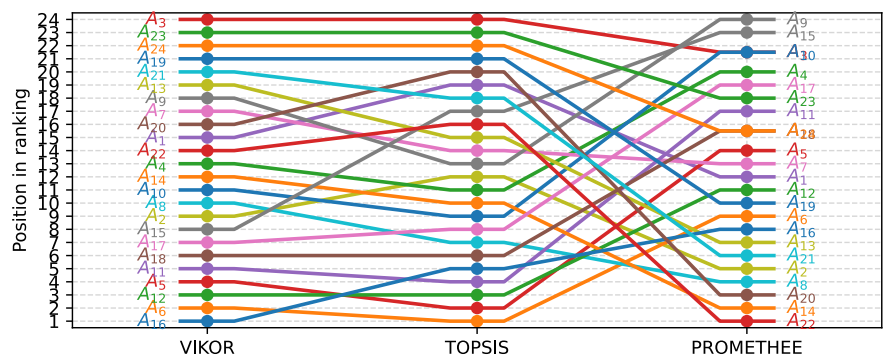

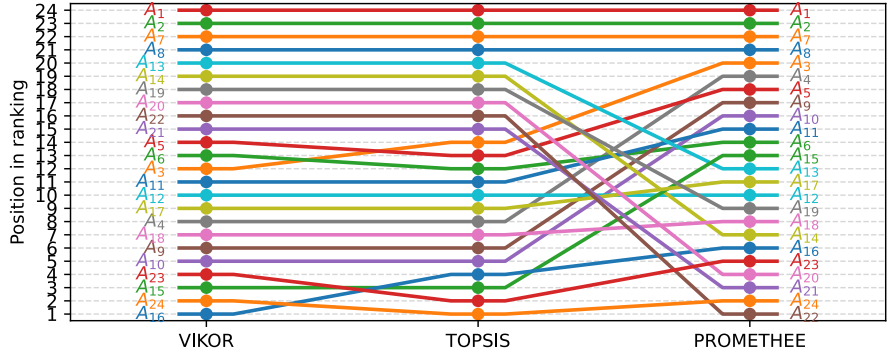

**Fig. 1.** Visualization of MCDM methods' ranking with Equal weight (top figure) and Entropy weight (bottom figure).

## 7 Conclusion

This paper explores the business feasibility of selling excess heat from an electrolyzer to district heating grids. By incorporating the sale of excess heat into three business models, this study assessed their impact on the ecosystem using agent-based modeling and three Multi-Criteria Decision-Making methods. The business models analyzed are constant hydrogen production; flexible production that adjusts to electricity prices; and hydrogen production using renewable energy sources, with and without excess heat sales.

The case study is Greenlab in Skive, Denmark, demonstrating that selling excess heat improves the economic viability of all models. Notably, the lowest hydrogen production cost is 2.795 EUR/kg under flexible operations linked to electricity spot prices, with excess heat sales and a fixed hydrogen price of 3.298 EUR/kg. However, achieving this cost required reducing hydrogen production by 12% compared to continuous operations.

In terms of environmental impact, producing hydrogen from grid electricity indirectly emits approximately 13,783.8 tons of $CO_2$ over four years starting in 2023. The most sustainable model proved to be hydrogen production from renewable sources combined with excess heat sales, priced at 3.5 EUR/kg. However, the operational feasibility of renewable sources is limited to 29% of potential hours due to weather conditions and energy capacity, leading to a competitive cost of 5.76 EUR/kg for green hydrogen in the case study.

This paper provides valuable insights into producing competitively priced, 100% green hydrogen using renewable energy facilities. Additionally, this paper outlines the environmental considerations associated with hydrogen production from grid electricity up to 2027.

### 7.1    Limitations and future work

The existing simulation model primarily addresses the Power-to-Hydrogen ecosystem, treating the district heating network as a straightforward heat demand and utilizing all accessible excess heat. It also simplifies the operation of the electrolyzer to basic on/off states, whereas in practice, it operates dynamically with an efficiency curve influenced by power input. Future studies aim to enhance the model by integrating the Power-to-Hydrogen process with the district heating network and incorporating dynamic electrolyzer operation. This will include adding the district heating business ecosystem to evaluate potential synergies between excess heat production and heating demand, potentially requiring adjustments in electrolyzer operations. Furthermore, relevant stakeholders within the electricity, hydrogen, and district heating market are being implemented to obtain the ecosystem dynamics and utilize the benefits of multi-agent-based simulation.

**Acknowledgments.** This research is part of the following projects: The project "INNOMISSION II: MissionGreenFuel- Digitalization and test for dynamic and flexible operation of PtX components and systems (DYNFLEX)" funded by Innova-tion Fund Denmark; IEA IETS Annex Task 21: part of the Project "Danish Participation in IEA IETS Task XXI - Decarbonizing industrial systems in a circular economy framework", funded by EUDP (project number: 134233-511205); IEA DHC Annex TS4: Digitalisation of District Heating and Cooling project, funded by EUDP (project number: 134-22011); Project "ClusterSoutH2 - Designing a PtX Ecosystem in Southern Denmark" funded by The European Regional Development Fund.

**Disclosure of Interests.** The authors have no competing interests to declare that are relevant to the content of this article.

## References

1.  A. J. Friedemann, "Why Not Electrify Commercial Transportation with Batteries?," in *Life after Fossil Fuels: A Reality Check on Alternative Energy*, A. J. Friedemann Ed. Cham: Springer International Publishing, 2021, pp. 41-45.
2.  S. Zhan, P. Hou, G. Yang, and J. Hu, "Distributionally robust chance-constrained flexibility planning for integrated energy system," *International Journal of Electrical Power & Energy Systems,* vol. 135, p. 107417, 2022/02/01/ 2022, doi: https://doi.org/10.1016/j.ijepes.2021.107417.
3.  C. Huang, Y. Zong, S. You, and C. Træholt, "Economic model predictive control for multi-energy system considering hydrogen-thermal-electric dynamics and waste heat recovery of MW-level alkaline electrolyzer," *Energy Conversion and Management,* vol. 265, p. 115697, 2022/08/01/ 2022, doi: https://doi.org/10.1016/j.enconman.2022.115697.
4.  W. Wang, C. Huang, and Y. Zong, "Model Predictive Control based Optimal Dispatch of Wind and Hydrogen Hybrid Systems," in *2021 International Conference on Advanced Technology of Electrical Engineering and Energy (ATEEE)*, 24-26 Dec. 2021 2021, pp. 74-80, doi: 10.1109/ATEEE54283.2021.00023.
5.  Fjernvarme Fyn. "Hvad betyder visningerne på måleren?" https://www.fjernvarmefyn.dk/viden/overblik/dit-fjernvarmeanlaeg/kend-din-maaler/din-maaler/hvad-betyder-visningerne-paa-maaleren (accessed December 07, 2022).

6. the Danish District Heating Association. "Power-to-X og fjernvarme." https://www.danskfjernvarme.dk/-/media/danskfjernvarme/gronenergi/analyser/ptx/fjernv-ptx-pixi.pdf (accessed December 07, 2022).
7. European Comission. "Hydrogen." https://energy.ec.europa.eu/topics/energy-systems-integration/hydrogen_en (accessed June 19, 2023).
8. L. Lück, P. Larscheid, A. Maaz, and A. Moser, "Economic potential of water electrolysis within future electricity markets," in *2017 14th International Conference on the European Energy Market (EEM)*, 6-9 June 2017 2017, pp. 1-6, doi: 10.1109/EEM.2017.7981950.
9. N. Fatras, Z. Ma, and B. N. Jørgensen, "A frequency-domain analysis of electricity market prices for multi-timescale flexibility of Power-to-X facilities," in *2022 18th International Conference on the European Energy Market (EEM)*, 13-15 Sept. 2022 2022, pp. 1-6, doi: 10.1109/EEM54602.2022.9921165.
10. M. Hermesmann and T. E. Müller, "Green, Turquoise, Blue, or Grey? Environmentally friendly Hydrogen Production in Transforming Energy Systems," *Progress in Energy and Combustion Science,* vol. 90, p. 100996, 2022/05/01/ 2022, doi: https://doi.org/10.1016/j.pecs.2022.100996.
11. Internation Renewable Energy Agency. "Hydrogen purchase agreements." https://www.irena.org/Innovation-landscape-for-smart-electrification/Power-to-hydrogen/15-Hydrogen-purchase-agreements (accessed.
12. T. D. Fechtenburg, "The value of flexibility for electrolyzers," Energinet,, 2022. [Online]. Available: https://energinet.dk/media/bonk4x1i/the-value-of-flexibility-for-electrolyzers-thomas-dalgas-fechtenburg-energinet.pdf
13. FfE, "Series of articles concerning hydrogen: Electrolyzer operating modes." [Online]. Available: https://www.ffe.de/en/publications/series-of-articles-concerning-hydrogen-electrolyzer-operating-modes/
14. Danish Ministry of Climate Energy and Utilities, "Stor interesse for PtX-udbud: Seks projekter får del i 1,25 milliarder kroner til dansk produktion af grøn brint," ed, 2023.
15. Z. Ma, B. N. Jørgensen, M. Levesque, M. Amazouz, and Z. Ma, "Business Models for Digitalization Enabled Energy Efficiency and Flexibility in Industry: A Survey with Nine Case Studies," in *Energy Informatics*, Cham, B. N. Jørgensen, L. C. P. da Silva, and Z. Ma, Eds., 2024// 2024: Springer Nature Switzerland, pp. 253-270.
16. K. Christensen, Z. Ma, Y. Demazeau, and B. N. Jørgensen, "Agent-based simulation framework for evaluating energy flexibility solutions and adoption strategies," in *1st Energy Informatics.Academy Asia Ph.D. Workshop, 28 May 2021*, Beijing, China, 2021: SpringerOpen Journal- Energy Informatics.
17. K. Christensen, Z. Ma, and B. N. Jørgensen, "Multi-agent Based Simulation for Investigating Electric Vehicle Adoption and Its Impacts on Electricity Distribution Grids and CO2 Emissions," in *Energy Informatics*, Cham, B. N. Jørgensen, L. C. P. da Silva, and Z. Ma, Eds., 2024// 2024: Springer Nature Switzerland, pp. 3-19.
18. M. Værbak, Z. Ma, Y. Demazeau, and B. N. Jørgensen, "A generic agent-based framework for modeling business ecosystems: a case study of electric vehicle home charging," *Energy Informatics,* vol. 4, no. 2, p. 28, 2021/09/24 2021, doi: 10.1186/s42162-021-00158-4.
19. AnyLogic. "AnyLogic webpage." https://www.anylogic.com/ (accessed April 11, 2024).
20. Z. Ma, "Business ecosystem modeling- the hybrid of system modeling and ecological modeling: an application of the smart grid," *Energy Informatics,* vol. 2, no. 1, p. 35, 2019/11/21 2019, doi: 10.1186/s42162-019-0100-4.
21. Z. Ma, K. Christensen, and B. N. Jorgensen, "Business ecosystem architecture development: a case study of Electric Vehicle home charging " *Energy Informatics,* vol. 4, p. 37, 24 June 2021 2021, Art no. 9 (2021), doi: https://doi.org/10.1186/s42162-021-00142-y.

Multi-agent based modeling for investigating excess heat utilization    13

22. K. Christensen, "Multi-Agent Based Simulation Framework for Evaluating Digital Energy Solutions and Adoption Strategies," ed: University of Southern Denmark, the Faculty of Engineering, 2022.
23. B. Kizielewicz, A. Shekhovtsov, and W. Sałabun, "pymcdm—The universal library for solving multi-criteria decision-making problems," *SoftwareX,* vol. 22, p. 101368, 2023/05/01/ 2023, doi: https://doi.org/10.1016/j.softx.2023.101368.
24. GreenLab. "GreenLab's webpage of their facilities." https://www.greenlab.dk/take-the-tour/ (accessed April 10, 2024).
25. Energinet. "Energi Data Service." https://www.energidataservice.dk/ (accessed May 17, 2023).
26. Ea Energianalyse, "Brint og PtX i fremtidens energisystem," 2020.
27. Danish Energy Agency. "Energistyrelsen bekræfter prisloft for overskudsvarme for 2023." https://ens.dk/presse/energistyrelsen-bekraefter-prisloft-overskudsvarme-2023 (accessed July 6, 2023).
28. Dansk Fjernvarme, Grøn Energi, COWI, and TVIS. "Power-to-x og fjernvarme." Dansk Fjernvarme. https://www.danskfjernvarme.dk/groen-energi/analyser/210512-power-to-x-og-fjernvarme (accessed January 02, 2022).
29. R. Saulnier, K. Minnich, and K. Sturgess, "Water for the Hydrogen Economy," 2020. [Online]. Available: https://watersmartsolutions.ca/wp-content/uploads/2020/12/Water-for-the-Hydrogen-Economy_WaterSMART-Whitepaper_November-2020.pdf
30. Danish Energy Agency, "Technology Data – Renewable fuels," 2017.
31. GreenLab Skive. "GreenLab's webpage of their facilities." https://www.greenlab.dk/take-the-tour/ (accessed June 16, 2023).
32. Danish Energy Agency, "Technology Data - Generation of Electricity and District Heating," 2023. [Online]. Available: https://ens.dk/sites/ens.dk/files/Analyser/technology_data_catalogue_for_el_and_dh.pdf
33. AnyLogic. "Optimization experiment information page." https://anylogic.help/anylogic/experiments/optimization.html (accessed April 11, 2024).
34. The Wind Power. "V136/4000-4200 - Power curve." https://www.thewindpower.net/turbine_en_1489_vestas_v136-4000-4200.php (accessed July 13, 2023).
35. EU Science Hub. "Introduction to solar radiation." European Comission. https://joint-research-centre.ec.europa.eu/photovoltaic-geographical-information-system-pvgis/getting-started-pvgis/pvgis-data-sources-calculation-methods_en (accessed August 08, 2023).
36. Clean Hydrogen Partnership. "Hydrogen cost and sales prices." https://h2v.eu/analysis/statistics/financing/hydrogen-cost-and-sales-prices (accessed August 30, 2023).